\documentclass[journal,12pt,onecolumn,draftclsnofoot]{IEEEtran}
\usepackage[T1]{fontenc}
\usepackage[latin9]{inputenc}
\usepackage{color}
\usepackage{textcomp}
\usepackage{amsmath}
\usepackage{amssymb}
\usepackage{graphicx}
\usepackage{caption}
\makeatletter
\providecommand{\tabularnewline}{\\}

\usepackage{cite}

\ifCLASSINFOpdf
\else
\fi
\usepackage{babel}

\usepackage{amsmath, amssymb, bbm, xspace}
\usepackage{epsfig}
\usepackage{longtable}
\usepackage{color}
\usepackage{mathrsfs}
\usepackage{comment}

\usepackage{courier}

\def\bkE{{\rm I\kern-.17em E}}
\def\bk1{{\rm 1\kern-.17em l}}
\def\bkD{{\rm I\kern-.17em D}}
\def\bkR{{\rm I\kern-.17em R}}
\def\bkP{{\rm I\kern-.17em P}}

\def\bkZ{{\bf{Z}}}

\def\bkE{{\rm I\kern-.17em E}}
\def\bk1{{\rm 1\kern-.17em l}}
\def\bkD{{\rm I\kern-.17em D}}
\def\bkR{{\rm I\kern-.17em R}}
\def\bkP{{\rm I\kern-.17em P}}

\makeatletter
\newcommand{\pushright}[1]{\ifmeasuring@#1\else\omit\hfill$\displaystyle#1$\fi\ignorespaces}
\newcommand{\pushleft}[1]{\ifmeasuring@#1\else\omit$\displaystyle#1$\hfill\fi\ignorespaces}
\makeatother

\def\bkZ{{\bf{Z}}}
\def\b12{(\beta_1,\beta_2)}

\newcounter{example}
\renewcommand{\theexample}{\thesection.\arabic{example}}

\newcounter{remark}
\renewcommand{\theremark}{\thesection.\arabic{remark}}

\def \lbf{{l}}
\def\Xscr{\mathcal{X}}

\def\Ebb{\mathbb{E}}
\newlength{\noteWidth}
\long\def\notes#1{\ifinner
{\tiny #1}
\else
\marginpar{\parbox[t]{\noteWidth}{\raggedright\tiny #1}}
\fi\typeout{#1}}

 \def\notes#1{\typeout{read notes: #1}} 

\newcommand{\eg}{e.g.\@\xspace} 

\newcommand{\Real}{\ensuremath{\mathbb{R}}}

\def\Ebb{\mathbb{E}}

\def\exp{\mathop{\hbox{\rm exp}}}

\def\spose#1{\hbox to 0pt{#1\hss}}

\def\text #1{\hbox{\quad#1\quad}}

\def\nthinsp{\mskip -2   mu}

\def\superstar{^{\raise 0.5pt\hbox{$\nthinsp *$}}}
\def\SUPERSTAR{^{\raise 0.5pt\hbox{$*$}}}

\def\lamstarT {\lambda^{\raise 0.5pt\hbox{$\nthinsp *$}T}}

\def\Lscr{{\cal L}}

\def\Xscr{{\cal X}}

\def\non{\nonumber}

\let\forallnew\forall
\renewcommand{\forall}{\forallnew\ }
\let\forall\forallnew

		\def\bkE{{\rm I\kern-.17em E}}
		\def\bk1{{\rm 1\kern-.17em l}}
		\def\bkD{{\rm I\kern-.17em D}}
		\def\bkR{{\rm I\kern-.17em R}}
		\def\bkP{{\rm I\kern-.17em P}}
		\def\bkY{{\bf \kern-.17em Y}}
		\def\bkZ{{\bf \kern-.17em Z}}
		\def\bkC{{\bf  \kern-.17em C}}

{\begin{list}{}%
         {\setlength{\leftmargin}{#1}}%
         \item[]%
}
{\end{list}}

		\def\bsp{\begin{split}}
		\def\beq{\begin{eqnarray}}
		\def\bal{\begin{align*}}
		\def\bc{\begin{center}}
		\def\be{\begin{enumerate}}
		\def\bi{\begin{itemize}}
		\def\bs{\begin{small}}
		\def\bS{\begin{slide}}
		\def\ec{\end{center}}
		\def\ee{\end{enumerate}}
		\def\ei{\end{itemize}}
		\def\es{\end{small}}
		\def\eS{\end{slide}}
		\def\eeq{\end{eqnarray}}
		\def\eal{\end{align*}}
		\def\esp{\end{split}}
		\def\qed{ \vrule height7.5pt width7.5pt depth0pt}  

	\def\cp2problem#1#2#3#4{\fbox
		 {\begin{tabular*}{0.9\textwidth}
			{@{}l@{\extracolsep{\fill}}l@{\extracolsep{6pt}}l@{\extracolsep{\fill}}c@{}}
				#1 & & $#4 $ 
			\end{tabular*}}}

		\def\bkE{{\rm I\kern-.17em E}}
		\def\bk1{{\rm 1\kern-.17em l}}
		\def\bkD{{\rm I\kern-.17em D}}
		\def\bkR{{\rm I\kern-.17em R}}
		\def\bkP{{\rm I\kern-.17em P}}
		
		\def\bkZ{{\bf{Z}}}

\newcommand {\beeq}[1]{\begin{equation}\label{#1}}
\newcommand {\eeeq}{\end{equation}}
\newcommand {\bea}{\begin{eqnarray}}
\newcommand {\eea}{\end{eqnarray}}

\def\texitem#1{\par\smallskip\noindent\hangindent 25pt
               \hbox to 25pt {\hss #1 ~}\ignorespaces}

\def\bsp{\begin{split}}
		\def\beq{\begin{eqnarray}}
		\def\bal{\begin{align*}}
		\def\bc{\begin{center}}
		\def\be{\begin{enumerate}}
		\def\bi{\begin{itemize}}
		\def\bs{\begin{small}}
		\def\bS{\begin{slide}}
		\def\ec{\end{center}}
		\def\ee{\end{enumerate}}
		\def\ei{\end{itemize}}
		\def\es{\end{small}}
		\def\eS{\end{slide}}
		\def\eeq{\end{eqnarray}}
		\def\eal{\end{align*}}
		\def\esp{\end{split}}
		\def\qed{ \vrule height7.5pt width7.5pt depth0pt}  

\usepackage{amsmath, amssymb, xspace}
\usepackage{epsfig}
\usepackage{longtable}
\usepackage{color}
\usepackage{mathrsfs}
\usepackage{subfig}

\usepackage{titlesec}
\titleformat{\subsection}[hang]{\normalsize\itshape}{\textup{\thesubsection}}{1em}{}[]

\makeatother

\usepackage{babel}
\begin{document}
\title{Free Energy Minimization: A Unified Framework for Modelling, Inference,
Learning, and Optimization 
}
\author{Sharu Theresa Jose and Osvaldo Simeone}

\maketitle
\vspace{-50pt}
The goal of these lecture notes is to review
the problem of free energy minimization as a unified framework underlying
the definition of maximum entropy modelling, generalized Bayesian
inference, learning with latent variables, statistical learning analysis of generalization,
and local optimization. Free energy minimization is first introduced,
here and historically, as a thermodynamic principle. Then, it is described
mathematically in the context of Fenchel duality. Finally, the mentioned
applications to modelling, inference, learning, and optimization are
covered starting from basic principles.\vspace{-15pt}

\subsection*{Relevance}

Free energy minimization is often invoked, implicitly or explicitly,
in different domains by taking an ad hoc approach that hides the generality
of the formulation and the common structure of the problem. Mathematical
details, e.g., in terms of measure theory, also often make some of
the material not easily accessible. These notes are intended to provide
an accessible reference for researchers interested in connecting the dots
among various standard problems in modelling, inference, learning,
and optimization within a common mathematical framework.\vspace{-15pt}

\subsection*{Prerequisites}

These notes require basic knowledge in probability and statistics.

\section*{Problem Statement}

The maximum entropy modelling principle, generalized Bayesian inference,
maximum likelihood learning with latent variables, Probably Approximately Correct (PAC) Bayes theory, mirror descent
optimization, as well as recent theories of human behavior, have all
in common their underlying reliance on an optimization principle first
enunciated at the end of the 19th century -- the minimization of
the free energy. Denoting as $\mathrm{E}_{\mathrm{x}\sim p(x)}[\cdot]$
the expectation operator for a random variable $\mathrm{x}\sim p(x)$,
the problem can be mathematically formulated as 
\begin{equation}
\textrm{\ensuremath{\underset{q}{\textrm{minimize}}} \ensuremath{J(q)=\underbrace{\mathrm{E}_{\mathrm{x}\sim q(x)}[L(\mathrm{x})]}_{\textrm{average loss (``energy'')}}+\hspace{0.2cm}T\times\underbrace{D(q)}_{\textrm{information ``complexity''}},} }\label{eq:minfe}
\end{equation}
where:\\
$ \noindent \bullet$  $q(\cdot)$, also denoted as $q$, is the probability distribution under optimization;\\
$\noindent \bullet$ $L(\cdot)$ is a loss function, which is also known as ``energy'' (or Hamiltonian in statistical physics); \\
$\noindent \bullet$ $T\geq0$ is a ``temperature'' parameter; and \\
$\noindent \bullet$ $D(q)$ is a convex penalty term, measuring the information-theoretic
``complexity'' of distribution $q(\cdot)$. 

The free energy $J(q)$ reflects a trade-off between performance,
as measured by loss function $L(\cdot)$, and  complexity, with the balance
between the two terms being dictated by the temperature level $T$.
The complexity penalty $D(q)$ may take different forms, depending
on the application. Notable examples are the Kullback-Leibler divergence $D(q)={\rm KL}(q||p)={\rm E}_{{\rm x} \sim q(x)}[\log (q({
\rm x})/p({\rm x}))]$ with respect to a reference distribution $p$, which represents a state of minimal complexity; and the negative entropy $D(q)=-{\rm H}(q)= {\rm E}_{{\rm x} \sim q(x)}[\log q({\rm x})]$. Informally, the information complexity $D(q)$ can be taken to measure  the amount of knowledge,
or information processing, needed to generate a sample $\mathrm{x}\sim q(x)$. For instance, with $D(q)=-{\rm H}(q)$, a small $D(q)$, and hence
a large entropy, amounts to a ``random'' pick, while a concentrated
distribution with a large $D(q)$ describes a more deliberate, and hence complex, choice. 

As we will discuss in these notes, the trade-off between performance and complexity measured by the free energy (\ref{eq:minfe}) plays a key role in modelling, inference, learning, and optimization problems. To see why, consider as an example the main problem in statistical learning of designing a learning algorithm that is able to generalizes well outside the training set. A learning algorithm that is too complex is bound to capture some of the noise in the training data that has no bearing on the true distribution of the data, causing overfitting and hence poor generalization. We can now associate the ``energy'' term in (1) with the performance of the algorithm on the training set, and the information ``complexity'' term as the fraction part of this performance that accounts for overfitting, being an artifact of the complexity of the algorithm. The real-world performance of the learning algorithm is then measured by the free energy (1).

The rest of these notes will expand on this discussion by tackling the following questions:\\
$\noindent \bullet$  \textit{What is the historical origin of problem (\ref{eq:minfe})?} (The answer
will briefly take us back to the 19th century.)\\
$\noindent \bullet$ \textit{What is the general form of the solution $q_{\mathrm{opt}}$ to problem (\ref{eq:minfe}) and its corresponding optimal value $J_{\mathrm{opt}}=J(q_\mathrm{opt})$?}
(The answer will involve Fenchel duality.)\\
$\noindent \bullet$ \textit{What are its applications to modelling, inference, learning, and optimization?}
(The answer will cover the maximum entropy principle, generalized
Bayesian inference, variational learning, PAC Bayes theory, and 
mirror descent.)

As some remark on notation, the $i$th element of a vector $a$ is denoted as $a_i$ or $[a]_i$. Given finite dimensional vectors $a=[a_i]$ and $b=[b_i]$ with the same number of elements, we use $a^T b=\sum_i a_i b_i$ to denote the dot product, with superscript $T$ indicating the transpose operator.  For a vector $a$, $||a||_2=\sqrt{a^Ta}=\sqrt{\sum_i a_i^2}$ is the $l_2$ norm, and $||a||_1=\sum_i |a_i|$ is the $l_1$ norm. We  write  $c^{+}=\max\{0,c\}$ for a scalar $c$. $1(\mathcal{E})$ denotes the indicator function that equals 1 when event $\mathcal{E}$ is true and zero otherwise. Finally, we denote random variables with Roman fonts and corresponding realizations with standard fonts.
\section{Origin}
The concept of free energy
goes back to the roots of thermodynamics, where it was introduced by von Helmholtz and Gibbs in the 1870s and 1880s to
measure the maximum amount of work that can be extracted from a closed thermodynamic
system at a constant temperature. Originally called ``available energy'', it captures the fact that only part of the energy of a system can be used for work.  If, for example, all
the molecules of a gas in a box move to the left, the corresponding  kinetic energy
can be used to drive a turbine. If, instead, the same kinetic energy is distributed
as random molecular motion, it cannot be fully transformed into work.
The maximum usable part of the energy is the (Helmholtz)
free energy
$  F=E -TS,$ where $E$ is the overall (kinetic) energy of the system; $T$ represents the system's temperature; and $S$ is the thermodynamic entropy. The entropy measures the level of ``complexity'' of the system in terms of the number of micro-states, \eg, the number of molecules' positions and momenta, that produce the given observed macro-state. 

\section{ Minimizing the Free Energy}
In this section, we will see
that problem (\ref{eq:minfe}) amounts to the computation of the Fenchel dual of the convex function $D(q)$, and that, as a consequence, solutions can be in principle easily
defined -- if not computed -- for a variety of information complexity measures $D(q).$ 

To start, we recall that, given a function $f(u),$ with $u$ being
a finite-dimensional real vector, its convex, or Fenchel, dual function $f^{*}(v)$ is
given as the solution of the problem
\begin{equation}
f^{*}(v)=\max_{u}\textrm{\ensuremath{u^T v }\ensuremath{-f(u)}}\label{eq:convex_dual},
\end{equation} 
where vector $v$ has the same dimension as $u$.
For a convex function $f(u)$, this can be interpreted as evaluating
the intercept $-f^{*}(v)$ for the tangent of slope $v$ of function $f(u)$ \cite{shalev2012online}. From now on, we will
assume that function $f(u)$ is convex, closed\footnote{$f(u)$ is closed if its epigraph, i.e., the set $\{(u,y):f(u)\leq y\}$,
is a closed set.}, and differentiable. From standard results in Fenchel duality, the optimal
solution $u_{\rm opt}(v)$ to problem (\ref{eq:convex_dual}) for any fixed vector $v$ is related to the dual function $f^{*}(v)$ by the relationship
\begin{equation}
u_{\mathrm{opt}}(v)=\nabla f^{*}(v).\label{eq:fenchel_dualoptimal}
\end{equation}
Furthermore, definition (\ref{eq:convex_dual}) implies the Fenchel-Young
inequality 
\begin{equation}
f^{*}(v)\geq\textrm{\ensuremath{u^Tv }\ensuremath{-f(u)}},\label{eq:FYineq}
\end{equation}
which holds for any vectors $u$ and $v$, 
with equality attained at $u=u_{\mathrm{opt}}(v)$. 

How is the free energy minimization
problem \eqref{eq:minfe} related to the computation \eqref{eq:convex_dual} of the Fenchel dual? To elaborate,
 assume that the alphabet $\mathcal{X}$
of random variable $\mathrm{x}$ in \eqref{eq:minfe} is discrete and finite, i.e., $\mathcal{X}=\{1,...,N_{x}\}$. We can then  write the negative free energy scaled by the temperature $T$ as \begin{align}
-\frac{J(q)}{T}=\mathrm{-E}_{\mathrm{x}\sim q(x)}\left[\frac{L(\mathrm{x})}{T}\right]-D(q)=\ensuremath{q^T\biggl(-\frac{\lbf}{T}\biggr) }\ensuremath{-D(q)},\label{eq:vector_minfe}
\end{align}
where vectors $q$ and $l$ are defined in Table~\ref{tab:FE-Fenchel}. Therefore, with the correspondence detailed in Table~\ref{tab:FE-Fenchel},
the maximization of \eqref{eq:vector_minfe}, and hence the minimization \eqref{eq:minfe}, is an instance of the Fenchel
duality problem \eqref{eq:convex_dual}. Note that in Table~\ref{tab:FE-Fenchel} the notation $I_{S}(q)=1(q \geq 0: ||q||_1=1)$
represents the support function for the set of all possible distributions $q$, i.e., the probability simplex. 
\begin{table}
\begin{center}
{\tiny{}}%
\begin{tabular}{|c|c|}
\hline 
Fenchel duality & free energy minimization\tabularnewline
\hline 
\hline 
$u$ & $q\doteq[q(1),...,q({N_{x}})]^{T}$\tabularnewline
\hline 
$v$ & $-l/T\doteq-[L(1),...,L(N_{x})]^{T}/T$ \tabularnewline
\hline 
$f(u)$ & $D(q)+I_{S}$$(q)$\tabularnewline
\hline 
$f^{*}(v)$ & $-J_{\mathrm{opt}}/T$\tabularnewline
\hline 
\end{tabular}{\tiny\par}
\vspace{-0.4cm}
\par\end{center}
\caption{Mapping between free energy minimization \eqref{eq:minfe} and Fenchel duality \eqref{eq:convex_dual}.} \label{tab:FE-Fenchel}
\end{table}
From \eqref{eq:fenchel_dualoptimal} and known Fenchel duality relationships, 
we obtain the solution of problem \eqref{eq:minfe} listed in Table~\ref{tab:FE-Dq} for the most common information complexity measures $D(q)$. We note that the conclusions in Table II apply also for a continuous alphabet, although the derivation does not follow directly from (\ref{eq:fenchel_dualoptimal}) and would require the use of calculus of variations. 
\begin{table}
\begin{center}
\begin{tabular}{|c|c|c|}
\hline 
$D(q)$ & minimum free energy $J_\mathrm{opt}$ & optimal solution $q_\mathrm{opt}(x)$ \tabularnewline
\hline 
\hline 
$-{\rm H(q)}$ & $-\log\sum_{x}\exp(-L(x)/T)$ & $\exp(-L(x)/T)/\sum_{x}\exp(-L(x)/T)$\tabularnewline
\hline 
KL$(q||p)$ & $-\log {\rm E}_{{\rm x} \sim p(x)}[\exp(-L({\rm x})/T)]$ & $p(x)\exp(-L(x)/T)/{\rm E}_{{\rm x} \sim p(x)}[\exp(-L({\rm x})/T)]$\tabularnewline
\hline 
$\frac{1}{2}||q-p||^{2}$ & $\textrm{mi\ensuremath{\textrm{n}_{q}}}||q-(p-l/T)||^{2}$ & $(p(x)-L(x)/T-\tau)^{+}$ \tabularnewline
\hline 
\end{tabular}
\vspace{-0.4cm}
\par\end{center}
\caption{Optimal solution of the free energy minimization problem \eqref{eq:minfe} for different choices of the information complexity $D(q)$. Parameter $\tau$ satisfies $\sum_x (p(x)-L(x)/T-\tau)^{+}=1$.
} \label{tab:FE-Dq}
\vspace{-0.8cm}
\end{table}
\begin{figure}
\begin{centering}
\includegraphics[scale=0.5,clip=true, trim = 0in  2.3in 0in 2.6in]{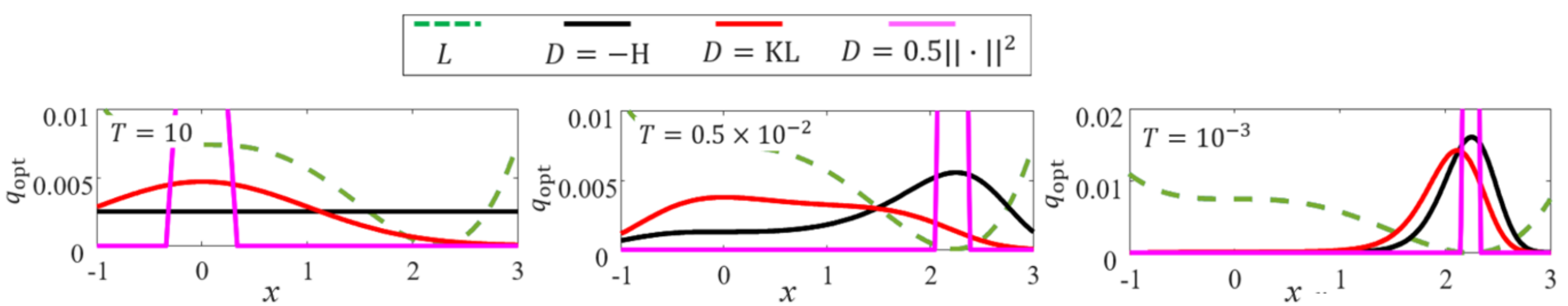}
\par\end{centering}
\caption{Loss function $L(x)$ and optimal solution $q_{\mathrm{opt}}$ of problem \eqref{eq:minfe} under the three complexity measures $D(q)=-{\rm H}(q)$, KL$(q||p)$ and $0.5 ||q-p||^2$ with prior distribution $p(x)=\mathcal{N}(x;0,1)$ as a function of $x$, for various values of the temperature $T$. }\label{fig:Fenchel}
\vspace{-0.6cm}
\end{figure}

As an example, Figure~\ref{fig:Fenchel} illustrates the trade-off between accuracy and complexity that results from free energy minimization when varying the temperature $T$ under the three complexity measures indicated in Table~\ref{tab:FE-Dq} with prior $p(x)=\mathcal{N}(x;0,1)$ being the zero-mean unit-variance Gaussian distribution. For $T=10$, the optimizing distribution $q_{\mathrm{opt}}$ puts more weight on minimizing the complexity term $D(q)$. As a result, for $D(q)=-{\rm H}(q)$, the optimal distribution  $q_{\mathrm{opt}}$ is close to a uniform distribution, whereas it approximates the prior $p$ for $D(q)=$  KL$(q||p)$  and $0.5 ||q-p||^2$. 
For smaller temperature values $T$, the emphasis shifts on minimizing $L(x)$ and the optimal distribution $q_{{\rm opt}}$ increasingly concentrates on the minimizer of the loss function.
\section{Applications}
\subsection*{\textcolor{black}{ $\mathrm{1}$. Maximum Entropy Principle }}
The maximum entropy principle provides a standard approach for the selection of a probabilistic model when the only information available on the quantity of interest consists of the averages of given statistics. Formally, suppose that we are given the expected values $\alpha_k$ of functions $f_k(x)$ for $k=1,\hdots,m$ of a random observation ${\rm x}$ with unknown distribution. The maximum entropy principle selects the probability distribution $q$ that solves the following optimization problem \cite{cover06elements}
\begin{align}
\max_{q} \quad & {\rm H}(q)\non \\
\mbox{s.t.} \quad &\mathrm{E}_{{\rm x} \sim q(x)}[f_k({\rm x})] = \alpha_k  \hspace{0.5cm}\mbox{for} \hspace{0.2cm} k=1,\hdots,m.  \label{eq:ME}
    \end{align}
Introducing a Lagrange multiplier $\lambda_k$ for each of the average constraints in \eqref{eq:ME}, the optimization of the Lagrangian for problem \eqref{eq:ME}  can be written as the free energy minimization (with $T=1$)
\begin{align}
\min_q J(q)=\mathrm{E}_{\mathrm{x}\sim q(x)}[\underbrace{-\sum_{k}\lambda_{k}(f_{k}(\mathrm{x}))}_{L(\mathrm{x})}]+\underbrace{(-{\rm H}(q))}_{D(q)}.
\end{align}
The solution of problem \eqref{eq:ME} then directly follows from Table~\ref{tab:FE-Dq} as
\begin{align}
q_\mathrm{opt}(x)\propto\exp\left(\sum_{k}\lambda_{k}f_{k}(x)\right). \label{eq:optimalq_ME}
\end{align}
The optimizing distribution in \eqref{eq:optimalq_ME} belongs to the exponential family with natural parameter vector $\lambda=(\lambda_1,\hdots,\lambda_m)$ and sufficient
statistics vector $f=(f_{1},\hdots,f_m)$ \cite{simeone2018brief}.

\subsection*{\textcolor{black}{$\mathrm{2}$.  Generalized Bayesian Inference}}
Consider a probabilistic model given by a joint distribution $p(x,y)$.
Bayesian inference of $\mathrm{x}$ given an observation $\mathrm{y}=y$
amounts to the computation of the posterior distribution $p(x|y)=p(x,y)/p(y)$.
This requires the evaluation of the partition function, or marginal likelihood,
$p(y)=\sum_{x}p(x,y)$ of the observation. More generally, given
an unnormalized distribution $\tilde{p}(x)$ -- \eg, $p(x,y)$ with
a fixed value of $y$ -- Bayesian inference requires the  computation of the normalizing partition function
$Z=\sum_{x}\tilde{p}(x)$. 

Define as the loss function $L(x)=-\log\tilde{p}(x)$ and as the information complexity penalty $D(q)=-\mathrm{H}(q)$. With this choice, by Table II, the minimum free energy with $T=1$ equals the negative log-partition function, i.e., $J_\mathrm{opt}=-\log Z=-\log \sum_{x}\tilde{p}(x)$ and the optimal distribution $q_\mathrm{opt}(x)$ equals the desired normalized distribution $p(x)=\tilde{p}(x)/Z$. Furthermore, by Table I and the Fenchel-Young inequality (\ref{eq:FYineq}), we obtain the following inequality, also known as \emph{Evidence
Lower BOund (ELBO)} \cite{simeone2018brief}
\begin{align}
\log Z & \geq -\mathrm{E}_{\mathrm{x}\sim q(x)}[\underbrace{-\log \tilde{p}(\mathrm{x})}_{L(\mathrm{x})}]+\underbrace{{\rm E}_{{\rm x} \sim q(x)}[-\log \ensuremath{q({\rm x})}]}_{\mathrm{H}(q)}=-J(q).\label{eq:ELBO}
\end{align}
To sum up, we have shown that: (\emph{i}) the log-partition function is
lower bounded by the negative free energy with the log-loss function
$L(x)=-\log \tilde{p}(\mathrm{x})$ and an entropy penalty;
and (\emph{ii}) the maximum of the lower bound is achieved for distribution $q(x)$
equal to the desired normalized distribution $p(x)$.


Returning to the problem of Bayesian inference, the derivation above
shows that the problem of computing the posterior $p(x|y)$ can be
equivalently framed as the minimization of the free energy $J(q)=\mathrm{E}_{\mathrm{x}\sim q(x)}[-\log p(\mathrm{x},y)]-{\rm H}(q)$.
When this optimization is not tractable, restricting the optimization
space to a parametric, or non-parametric, family of distributions
yields an approximate Bayesian inference approach known as \textit{variational
inference}, which is central to state-of-the-art scalable solutions to Bayesian
inference \cite{simeone2018brief, mackay2003information,angelino2016patterns}. Furthermore, allowing for a larger class of divergence penalty terms $D(q)$, for
any loss function $L(x)$ -- not necessarily the log-loss -- and for any value
of the temperature $T>0$ yields the framework of \emph{generalized Bayesian
inference} \cite{knoblauch2019generalized}.
We finally note that 
the ELBO (\ref{eq:ELBO}) is also central
to Bayesian and Minimum Description Length
(MDL) model selection strategies \cite{grunwald2005advances,mackay2003information}.

\subsection*{\textcolor{black}{$\mathrm{3}$. Parametric Learning with Latent Variables}}

Consider now the problem of learning a probabilistic model in the
presence of latent, or unobserved, variables. The probabilistic model
is defined by a joint distribution $p_{\theta}(x,y)$ over observations
$\mathrm{y}$ and latent variables $\mathrm{x}$ that is parameterized by
a vector $\theta$. Maximum likelihood learning is based on the maximization
of the marginal log-likelihood $\log p_{\theta}(y)=\log \sum_{x}p_{\theta}(x,y)$ for an observation ${\rm y}=y.$
Denoting the ELBO (\ref{eq:ELBO}) as $J_{\theta}(q)=\mathrm{E}_{\mathrm{x}\sim q(x)}[-\log p_{\theta}(\mathrm{x},y)]-{\rm H}(q)$ in order to highlight its dependence on the model parameter vector $\theta$, this problem can be reformulated as the minimax optimization\begin{equation}
\textrm{\ensuremath{\underset{\theta}{\textrm{min}}} }\underset{q} {\textrm{max}} \hspace{0.3cm}J_{\theta}(q).\label{eq:EM}
\end{equation}
This is because, as we have discussed in the previous section, the
inner maximization yields the marginal log-likelihood $\log p_{\theta}(y)$
with the optimum achieved for $q(x)=p_{\theta}(x|y)$. 

The Expectation Maximization (EM) algorithm tackles the ML problem
(\ref{eq:EM}) by iteratively optimizing over distribution $q$ for
fixed $\theta$ -- which amounts to the Bayesian inference problem
covered in the previous section -- and optimizing over the parameter
vector $\theta$ for a fixed $q$. In modern applications, neither of the two steps is typically tractable, and approximate solutions are used in which the inner maximization is replaced by variational inference and stochastic gradient-based descent-ascent methods are implemented \cite{angelino2016patterns}.
Variants that consider alternative loss measures and the divergence metrics include the variational information bottleneck approach \cite{alemi2016deep}.

\subsection*{\textcolor{black}{$\mathrm{4}$. Statistical Learning Theory}}
Assume that we have a training data set
 ${\rm s}=({\rm z}_1,\hdots, \rm{z}_m)$ of $m$ examples generated independently as ${\rm z}_i \sim p(z)$, with $i=1,\hdots,m,$ from some unknown distribution $p(z)$.
A (probabilistic) learning algorithm is described by a stochastic mapping $q(x|s)$ between the training data set and a model parameter ${\rm x}$. The algorithm is said to generalize if it ``performs well" on a new, independently generated test example ${\rm z}'\sim p(z)$. Denoting as $\ell(x,z)$ the loss accrued by model $x$ when applied to data $z$, the performance criterion of interest for a model parameter $x$ is hence the test loss  $\mathcal{L}( x)=\mathrm{E}_{{\rm z'} \sim p(z)}[\ell(x,{\rm z}')]$  \cite{simeone2018brief, mackay2003information,angelino2016patterns}. For a given training set ${\rm s}=s$, the learner, however, has only access to the training loss $\mathcal{L}_{s}( x)=\frac{1}{m} \sum_{i=1}^m \ell(x,z_i)$. The difference between the test loss and the training loss $\Delta \mathcal{L}_{s}({ x})=\mathcal{L}({ x})-\mathcal{L}_{ s}({x})$ is known as the \emph{generalization gap} or generalization error. In the case of a probabilistic algorithm, one is typically interested in its average $\Delta \mathcal{L}_s = {\rm E}_{{\rm x}\sim q(x|s)}[\Delta \mathcal{L}_{s}({\rm x})]$. This is an important metric to gauge the generalization capacity of a training procedure $q(x|s)$. In fact, if the generalization gap is small, a low training loss, which can be guaranteed through the design of the training procedure $q(x|s)$, implies a low test loss.

While the average generalization gap $\Delta \mathcal{L}_s$ cannot be evaluated due to its dependence on the unknown test loss, one of the main concerns of statistical learning theory  is obtaining computable upper bounds on it. As we will discuss, these bounds can be used both to define learning criteria and to obtain generalization measures. 

To see how the framework of free energy minimization can be useful for this purpose, define as the loss function $L(x)=-\beta \Delta \mathcal{L}_s(x)$ and as the information complexity penalty $D(q)=\mathrm{KL}(q(x|s)||p(x))$ for some constant $\beta>0$ and prior distribution $p(x)$. With this choice, by Table II, the minimum free energy with $T=1$ is given as $J_\mathrm{opt}=-\log{\rm E}_{{\rm x}\sim p(x)}[\exp(\beta\Delta \mathcal{L}_{{s}}({\rm x}))]$. This is the negative \emph{log-moment generating function}, also known as negative cumulant generating function, of random variable $\Delta \mathcal{L}_s({\rm x})$ under the prior distribution ${\rm x}\sim p(x)$. Furthermore, by Table I and the Fenchel-Young inequality (\ref{eq:FYineq}), we obtain the following upper bound -- an instance of the \emph{Donsker-Varadhan inequality} -- on the average generalization gap\begin{align}
    {\rm E}_{{\rm x}\sim q(x|s)}[\beta\Delta \mathcal{L}_{{s}}({\rm x})]  \leq {\rm KL}(q(x|s)||p(x)) + \log {\rm E}_{{\rm x}\sim p(x)}[\exp(\beta\Delta \mathcal{L}_{{ s}}({\rm x}))]. \label{eq:DV_gen}
\end{align}  

Let us now use the bound (\ref{eq:DV_gen}) to define a learning criterion. To this end, we note that this inequality directly implies that the average test loss $\mathrm{E}_{{\rm x} \sim q(x|s)}[\Lscr({\rm x}) ]$ -- the true objective of a learning procedure -- can be upper bounded by the regularized (average) training loss $\mathrm{E}_{{\rm x} \sim q(x|s)}[\Lscr_s({\rm x}) ]+\beta^{-1} {\rm KL}(q(x|s)||p(x))$ plus an additional term -- the log-moment generating function -- that does not depend on the learning algorithm $q(x|s)$ under optimization. Hence, for a given prior distribution $p(x)$ and $\beta>0$, a learning algorithm $q(x|s)$ that minimizes an upper bound on the average test loss is obtained by addressing the free energy minimization problem 
\begin{align}
\min_{q(x|s)}  \mathrm{E}_{{\rm x} \sim q(x|s)}[\mathcal{L}_{ s}({\rm x})]+ \frac{1}{\beta} {\rm KL}(q(x|s)||p(x)). \label{eq:IRM}
\end{align}This approach is known as \emph{information risk minimization}  \cite{zhang2006information}, and it can be seen to be a special case of generalized Bayesian inference (see Sec. II.2). The optimal choice of  the learning algorithm  follows from Table~\ref{tab:FE-Dq} as the Gibbs distribution $q_{{\rm opt}}(x|s) \propto p(x)\exp(-\beta \mathcal{L}_{s}({\rm x})) $. 


The inequality \eqref{eq:DV_gen} can also be used to obtain an explicit upper bound on the generalization gap. Specifically, PAC Bayes theory seeks bounds that hold with high probability with respect to ${\rm s} \sim p(s)$ for any data distribution $p(s)$. This derivation is made difficult by the presence of the log-moment generating function (second term on the right-hand side in \eqref{eq:DV_gen}), which depends on the unknown test loss. To proceed, one needs to make additional assumptions on the distribution of the loss function. The most basic derivation of PAC Bayes bounds assumes that the loss function is bounded, i.e., $a\leq l(\cdot,\cdot)\leq b$ for some parameters $0\leq a<b$. In this case, an application of the Markov and Hoeffding inequalities followed by an optimization over $\beta$ yields the following result 
\cite{guedj2019still}: With probability at least $1-\delta$ over the training data ${\rm s} \sim p(s)$ for any distribution $p(s)$, the following upper bound on the average generalization gap holds
\begin{equation}   \mathrm{E}_{{\rm x} \sim q(x|s)}[ \Delta \mathcal{L}_{\rm s}({\rm x})] \leq \sqrt{\frac{(b-a)^2}{2m}\biggl({\rm KL}(q(x|s)||p(x))+\log \frac{1}{\delta}\biggr)}.
\end{equation}

\subsection*{\textcolor{black}{$\mathrm{5}$. Estimation of Information-Theoretic Metrics}}
The estimation from data of information-theoretic metrics, such as entropy and mutual information, is a crucial step in many machine learning and data science applications.  Information-theoretic metrics can generally be expressed in terms of the KL divergence. For instance, the mutual information $I({\rm{x}};\rm{y})$ between two random variables ($\rm{x},\rm{y}$) jointly distributed according to $p(x,y)$ can be written as the divergence $I(\rm{x};\rm{y}) =$ KL$(p(x,y)||p(x)p(y))$ between the joint distribution and the product of its marginals. Therefore, the estimation of information-theoretic metrics often relies on the estimate of the KL divergence  KL$(p(x)||q(x))$ from data samples drawn from the two distributions $p(x)$ and $q(x)$.

The free-energy minimization principle, in the form of Donsker-Varadhan inequality (cf. (\ref{eq:DV_gen})), is a key tool to estimate the KL divergence from data. To see this, by
setting $D(q)=$ KL$(p(x)||q(x))$ and $T=1$, Table~\ref{tab:FE-Dq} gives the Donsker-Varadhan inequality
\begin{align}
{\rm{KL}}(p(x)||q(x)) \geq \Ebb_{{\rm{x}} \sim p(x)}[-L({\rm{x}})]-\log \Ebb_{{\rm{x}}\sim q(x)}[\exp(-L({\rm{x}}))], \label{eq:KLbound}
\end{align} which holds for any loss function $L(x)$. In particular, the relation \eqref{eq:KLbound} holds with equality for the optimizing loss function $L^{*}(x)=-\log (p(x)/q(x))$, yielding the following variational form of the KL divergence 
\begin{align}
{\rm{KL}}(p(x)||q(x))=\sup_{L(\cdot) \in \mathcal{L}} \{ \Ebb_{{\rm{x}} \sim p(x)}[-L({\rm{x}})]-\log \Ebb_{{\rm{x}}\sim q(x)}[\exp(-L({\rm{x}}))]\}, \label{eq:optimization}
\end{align} where $\mathcal{L}$ is the space of real-valued functions $L:\Xscr \rightarrow \Real$. A practical estimator can now be obtained by (\emph{i}) replacing the two expectations in  \eqref{eq:optimization} with empirical averages over the data sets of samples drawn from distributions $p(x)$ and $q(x)$, respectively; and (\emph{ii}) optimizing over a tractable subset of functions $L(\cdot)$ such as neural networks. This is the approach taken by the Mutual Information Neural Estimator (MINE) \cite{belghazi2018mine}.

\subsection*{\textcolor{black}{$\mathrm{6}$. Local Optimization}}

The standard gradient descent algorithm for a differentiable function
$g(x)$ produces a sequence of iterates $x^{(i)},$ for
$i=1,2,...$, with the $(i+1)$th iterate obtained as 
\begin{equation}
\textrm{\ensuremath{x^{(i+1)}=\arg\underset{x}{\min}} \hspace{0.2cm}\ensuremath{}\ensuremath{\nabla g(x^{(i)})^{T}x+\frac{1}{2\alpha^{(i)}}||x-x^{(i)}||^{2},}}\label{eq:gradient}
\end{equation}
where $\alpha^{(i)}>0$ is a step size. The $(i+1)$th iterate in \eqref{eq:gradient} is
hence the minimum of a strongly convex approximation of function $g(x)$
that has the same gradient $\nabla g(x^{(i)})$ at the current iterate
$x^{(i)}.$ The quadratic term $||x-x^{(i)}||^{2}$ penalizes 
deviations from the current iterate by an amount that depends on the
inverse of the step size. This ensures that the next iterate is localized within a ``trust region'' defined by the current iterate. It can be easily seen that the solution
of problem (\ref{eq:gradient}) yields the familiar update $x^{(i+1)}=x^{(i)}-\alpha^{(i)}\nabla g(x^{(i)})$. 

By (\ref{eq:gradient}), gradient descent penalizes deviations from the current
iterate by assuming that the relevant geometry is Euclidean so that
the distance is measured by the term $||x-x^{(i)}||^{2}$. In problems
involving the optimization over a probability distribution $q$, the
Euclidean space, and associated distance metric $||x-x^{(i)}||^{2}$,
are no longer natural choices, and other ``distance'' metrics, such
as the KL divergence, have a more direct relevance.

Based on the above, considering a discrete distribution $q=[q(1), \hdots q(N_x)]^T$, one can define a variant of gradient descent
that addresses, at each iteration $i$, the problem
\begin{equation}
\textrm{\ensuremath{q^{(i+1)}=\arg \underset{q}{\rm{min }}}\ensuremath{}\hspace{0.2cm}\ensuremath{\nabla g(q^{(i)})^{T}q+\frac{1}{\alpha^{(i)}}\biggl(\textrm{KL}(q||q^{(i)})1_{S}(q)}}\biggr).\label{eq:gradient-1}
\end{equation}
This is an instance of the minimization (\ref{eq:minfe})
of the free energy with $L(x)=[\nabla g(q^{(i)})]_x$, $T=1/\alpha^{(i)}$, and $D(q)=\textrm{KL}(q||q^{(i)})1_{S}(q)$.
It is also an instance of \emph{mirror descent}, which, more generally, tackles a problem in the form (\ref{eq:gradient-1}) with a Bregman divergence as penalty and minimization constrained within a convex set \cite{beck2017first}. By Table~\ref{tab:FE-Dq}, the solution to problem \eqref{eq:gradient-1} is given as 
\begin{align}
q^{(i+1)}(x)\propto q^{(i)}(x) \exp(-\alpha^{(i)}[\nabla g(q^{(i)})]_x).\label{eq:-1}
\end{align}
yielding the Normalized Exponentiated Gradient algorithm \cite{shalev2012online}.

\section*{What We Have Learned}
The minimization of the free energy is a general principle that underlies many of the technique that are commonly used in signal processing and machine learning for the key tasks of model selection, inference, learning from data, and local optimization. These notes have aimed at elucidating this unifying thread in simple terms by starting from first principles via Fenchel duality. More connections not covered here include the use of the Donsker-Varadhan inequality (\ref{eq:DV_gen}) for the estimation of divergences and mutual information from data (see \cite{pantazis2020cumulant} and references therein) and entropy-regularized optimal transport.
\section*{{\small{}{{AUTHORS}}}}

\textbf{Sharu Theresa Jose} (sharu.jose@kcl.ac.uk) is a Postdoctoral Research Associate with the Department of Engineering at King's College London. She received her Ph.D. from the Systems and Control Engineering Department of Indian Institute of Technology, Bombay in 2018. Her research interests include information theory, statistical learning theory, stochastic control and optimization. 

\textbf{Osvaldo Simeone} (osvaldo.simeone@kcl.ac.uk) is a Professor of Information Engineering
with the Centre for Telecommunications Research at the Department
of Informatics of King's College London, where he directs King's Communications, Learning, and Information Processing (KCLIP) lab. He is a Fellow
of the IEEE.
\bibliographystyle{ieeetr}
\bibliography{ref}

\begin{thebibliography}{10}

\bibitem{shalev2012online}
S.~Shalev-Shwartz {\em et~al.}, ``Online {L}earning and {O}nline {C}onvex
  {O}ptimization,'' {\em Foundations and Trends{\textregistered} in Machine
  Learning}, vol.~4, no.~2, pp.~107--194, 2012.

\bibitem{cover06elements}
T.~M. Cover and J.~A. Thomas, {\em Elements of Information Theory, 2nd
  Edition,}.
\newblock {Wiley-Interscience}, July 2006.

\bibitem{simeone2018brief}
O.~Simeone, ``A {B}rief {I}ntroduction to {M}achine {L}earning for
  {E}ngineers,'' {\em Foundations and Trends{\textregistered} in Signal
  Processing}, vol.~12, no.~3-4, pp.~200--431, 2018.

\bibitem{mackay2003information}
D.~J. MacKay, {\em {I}nformation theory, {I}nference and {L}earning
  {A}lgorithms}.
\newblock Cambridge {U}niversity {P}ress, 2003.

\bibitem{angelino2016patterns}
E.~Angelino, M.~J. Johnson, and R.~P. Adams, ``{P}atterns of {S}calable
  {B}ayesian {I}nference,'' {\em Foundations and Trends{\textregistered} in
  Signal Processing}, vol.~9, pp.~119--247, Nov 2016.

\bibitem{knoblauch2019generalized}
J.~Knoblauch, J.~Jewson, and T.~Damoulas, ``Generalized {V}ariational
  {I}nference,'' {\em arXiv preprint arXiv:1904.02063}, 2019.

\bibitem{grunwald2005advances}
P.~D. Gr{\"u}nwald, I.~J. Myung, and M.~A. Pitt, {\em Advances in {M}inimum
  {D}escription {L}ength: {T}heory and {A}pplications}.
\newblock MIT {P}ress, 2005.

\bibitem{alemi2016deep}
A.~A. Alemi, I.~Fischer, J.~V. Dillon, and K.~Murphy, ``Deep {V}ariational
  {I}nformation {B}ottleneck,'' {\em arXiv preprint arXiv:1612.00410}, 2016.

\bibitem{zhang2006information}
T.~Zhang, ``Information-{T}heoretic {U}pper and {L}ower {B}ounds for
  {S}tatistical {E}stimation,'' {\em IEEE Transactions on Information Theory},
  vol.~52, no.~4, pp.~1307--1321, 2006.

\bibitem{guedj2019still}
B.~Guedj and L.~Pujol, ``Still {N}o {F}ree {L}unches: {T}he {P}rice to {P}ay
  for {T}ighter {PAC-B}ayes {B}ounds,'' {\em arXiv preprint arXiv:1910.04460},
  2019.

\bibitem{belghazi2018mine}
M.~I. Belghazi, A.~Baratin, S.~Rajeswar, S.~Ozair, Y.~Bengio, A.~Courville, and
  R.~D. Hjelm, ``{MINE}: {M}utual {I}nformation {N}eural {E}stimation,'' {\em
  arXiv preprint arXiv:1801.04062}, 2018.

\bibitem{beck2017first}
A.~Beck, {\em First-{O}rder {M}ethods in {O}ptimization}, vol.~25.
\newblock SIAM, 2017.

\bibitem{pantazis2020cumulant}
Y.~Pantazis, D.~Paul, M.~Fasoulakis, Y.~Stylianou, and M.~Katsoulakis,
  ``Cumulant {GAN},'' {\em arXiv preprint arXiv:2006.06625}, 2020.

\end{thebibliography}

\end{document}